\documentclass[aps,prd,superscriptaddress,preprintnumbers]{revtex4}
\usepackage{graphicx,float,wrapfig,subfigure}
\usepackage{amsfonts,amsmath,amssymb,amstext}
\usepackage{latexsym}
\usepackage{bm}
\usepackage{color}
\usepackage[normalem]{ulem}

\newcommand{\be}{\begin{equation}}
\newcommand{\ee}{\end{equation}}
\newcommand{\ba}{\begin{eqnarray}}
\newcommand{\ea}{\end{eqnarray}}

\newcommand{\non}{\nonumber\\}
\definecolor{red}{rgb}{0.7,0,0}
\definecolor{green}{rgb}{0,0.5,0}

\begin{document}

\title{(pseudo)Scalar mesons in a self-consistent NJL model}
\date{\today}
\author{Xiaozhu Yu}
\affiliation{Department of Physics, Jiangsu University, Zhenjiang 212013, P.R. China}
\author{Xinyang Wang}
\email{wangxy@ujs.edu.cn}
\affiliation{Department of Physics, Jiangsu University, Zhenjiang 212013, P.R. China}
\affiliation{School of Fundamental Physics and Mathematical Sciences, Hangzhou Institute for Advanced Study, UCAS, Hangzhou 310024, P.R. China}

\begin{abstract}
In this study, we investigate the mass spectrum of $\pi$ and $\sigma$ mesons at finite chemical potential using the self-consistent NJL model and the Fierz-transformed interaction Lagrangian. The model introduces an arbitrary parameter $\alpha$ to reflect the weights of the Fierz-transformed interaction channels. We show that when $\alpha$ exceeds a certain threshold value, the chiral phase transition transforms from a first-order one to a smooth crossover, which is evident from the behaviors of the chiral condensates and meson masses. Additionally, at high chemical potential, the smaller the value of $\alpha$, the higher the masses of the $\pi$ and $\sigma$ mesons become. Moreover, the Mott and dissociation chemical potentials both increase with the increase in $\alpha$. Thus, the meson mass emerges as a valuable experimental observable for determining the value of $\alpha$ and investigating the properties of the chiral phase transition in dense QCD matter.
\end{abstract}
\maketitle
\section{Introduction}
Exploring the properties of strongly interacting matter is a fundamental question in high energy nuclear physics. In particular, from a theoretical point of view, it is very important to study the breaking and restoration of the chiral symmetry in order to understand such kind of matter. As we know, the (approximate) chiral symmetry is believed to be a good symmetry in the light quark sector. Unfortunately, the perturbative method becomes unavailable for quantum chromodynamics (QCD) in low energy regime, since the strong coupling constant is no longer small enough. Besides, the lattice QCD can not handle the numerical calculations at finite chemical potential because of the famous sign problem. Hence, the effective theories and models are needed to investigate the QCD matter. Especially, based on the chiral symmetry and chiral symmetry breaking of QCD, the Nambu-Jona-Lasinio (NJL) model~\cite{Nambu:1961tp,Nambu:1961fr} is one of the most useful tools to study the properties of strongly interacting matter, such as the dynamical breaking/restoration of the chiral symmetry and the masses of light mesons.

One of the uncertainties of the NJL model is the way of dealing with the mean field approximation, and this issue has been well emphasized for a few decades~\cite{Klevansky:1992qe}. Mathematically, the Fierz transform of NJL model Lagrangian should be of equal importance, as compared with the original NJL model Lagrangian, but they are treated unequally when applying the mean field approximation. Therefore, we rewrite the Lagrangian as $\mathcal{L}_{R} = (1-\alpha)\mathcal{L}+\alpha \mathcal{L}_F$ by introducing an arbitrary weighting parameter $\alpha$, where $\mathcal{L}$ is the original NJL Lagrangian and $\mathcal{L}_F$ is the Fierz transform of $\mathcal{L}$. It has been discussed that there are no physical requirements for the choice of $\alpha$ value. And the value of $\alpha$ could be determined by astronomy observations, i.e., the properties of compact stars~\cite{Wang:2019uwl,Zhao:2019xqy,Wang:2019jze,Wang:2019npj} which impose constraints on the QCD equations of state.

In this paper, we will discuss a possible alternative way of predicting the value of $\alpha$ by the properties of light mesons. As we know, the mass spectra of pseudoscalar meson $\pi$ and scalar meson $\sigma$ have been studied in the NJL type model since a few dozen years ago~\cite{Asakawa:1989bq}. Since $\pi$ and $\sigma$ mesons are chiral partners, the mass difference between them carries the information of the chiral symmetry breaking and restoration. Therefore, apart from the indirect measurement on the equation of states of compact stars, the measurement of the meson masses in the heavy ion collision experiments is an alternative method to extract the information of the weighting parameter $\alpha$, as well as that of the chiral phase transition in dense QCD matter.

This paper is organized as follows. We begin with the general formalism in Sec.~\ref{sec-2}, and then the corresponding numerical results are presented in Sec.~\ref{sec-3}. Finally, the conclusions are given in Sec.~\ref{sec-4}.

\section{Formalism}
\label{sec-2}
The redefined Lagrangian---the combination of the original Lagrangian $\mathcal{L}$ in the two-flavor NJL model and the corresponding Fierz transformed Lagrangian $\mathcal{L}_F$ is given by~\cite{Wang:2019uwl}:
\ba
\mathcal{L}_{R} = (1-\alpha)\mathcal{L}+\alpha \mathcal{L}_F,
\ea
where
\be
\mathcal{L}= \bar{\psi}(i \partial\!\!\!/-m) \psi+ G\left[(\bar{\psi} \psi)^{2}+\left(\bar{\psi} i \gamma^{5} \cdot \tau \psi\right)^{2}\right]
\ee
and
\ba
\mathcal{L}_F&=& \bar{\psi}(i \partial\!\!\!/-m) \psi+
\frac{G}{8 N_{c}}\left[2(\bar{\psi} \psi)^{2}+2\left(\bar{\psi} i \gamma^{5} \tau \psi\right)^{2}-2(\psi \tau \psi)^{2}\right.
-2\left(\bar{\psi} i \gamma^{5} \psi\right)^{2}\nonumber\\
&&-4\left(\bar{\psi} \gamma^{\mu} \psi\right)^{2}-4\left(\bar{\psi} i \gamma^{\mu} \gamma^{5} \psi\right)^{2}
+\left.\left(\bar{\psi} \sigma^{2 m} \psi\right)^{2}-\left(\bar{\psi} \sigma^{\mu \nu} \tau \psi\right)^{2}\right],
\ea
along with the quark current masses $m = diag(m_u,m_d)$.

By applying the mean field approximation on the Lagrangian and dropping the irrelevant part, we get the effective Lagrangian
\ba
\left<\mathcal{L}_R\right>_{eff}&=&\bar{\psi}(i \partial\!\!\!/-M)\psi+G\left(1-\alpha+\frac{\alpha}{4N_c}\right)\sigma^2+\frac{\alpha G}{2 N_c}n^2,
\ea
where $\sigma = \left<\bar{\psi}\psi\right>$ is the quark condensation, and the quark number density is $n = \left<\psi^{\dagger}\psi\right>$. Besides, we have introduced the constituent quark mass $M = m - 2G\left(1-\alpha+\frac{\alpha}{4N_c}\right)\sigma$ and the renormalized chemical potential $\mu_r = \mu -\frac{\alpha G}{N_c}n$.

Thus, the corresponding thermodynamic potential density takes the form
\ba
\Omega &=& -\frac{T}{V}\ln Z \non
&=& G\left(1-\alpha+\frac{\alpha}{4N_c}\right)\sigma^2 - \frac{\alpha G}{2 N_c} n^2\non &&
-\frac{N_c N_f}{\pi^2}\int_{0}^{\Lambda}dp p^2\left\{E(M,p)+T \ln\left[1+\exp\left(-\frac{E(M,p)+\mu_r}{T}\right)\right] +T \ln\left[1+\exp\left(-\frac{E(M,p)-\mu_r}{T}\right)\right] \right\}.
\ea
Here, the energy dispersion relation $E(M,p)=\sqrt{M^2+p^2}$ and the Fermi-Dirac distribution functions
\ba
n(p,\mu)=\frac{1}{1+\exp\left(\frac{E(M,p)-\mu}{T}\right)}, ~~~~~~~~   \bar{n}(p,\mu)=\frac{1}{1+\exp\left(\frac{E(M,p)+\mu}{T}\right)}.
\ea

Then, the gap equations are determined by$\frac{\partial \Omega}{\partial M} = \frac{\partial \Omega}{\partial \mu_r} = 0$, which could be written into the exact form
\begin{subequations}
\label{Gap-quark}
\ba
\sigma + \frac{N_c N_f M}{\pi^2}\int_{0}^{\Lambda}\frac{dp p^2}{E(M,p)} \left[1- n(p,\mu_r)-\bar{n}(p,\mu_r)\right]=0
\ea
and
\ba
n- \frac{N_c N_f M}{\pi^2}\int_{0}^{\Lambda}\frac{dp p^2}{E(M,p)} \left[ n(p,\mu_r)-\bar{n}(p,\mu_r)\right]=0.
\ea
\end{subequations}
On the other hand, in order to calculate the meson masses, we obtain the dispersion relations for $\pi$ and $\sigma$ mesons in the random-phase approximation (RPA),
\begin{subequations}
\ba
\label{GapEq-pion}
1 - 2\left(1-\alpha+\frac{\alpha}{4N_c}\right)G\frac{N_c N_f}{\pi^2}P\int_{0}^{\Lambda}\frac{p^2}{E(M,p)}\left(1-\frac{M_{\pi}^2}{M_{\pi}^2-4E(M,p)^2}\right)\left(1- n(p,\mu_r)- \bar{n}(p,\mu_r)\right)dp = 0,
\ea
and
\ba
\label{GapEq-sigma}
1 - 2\left(1-\alpha+\frac{\alpha}{4N_c}\right)G\frac{N_c N_f}{\pi^2}P\int_{0}^{\Lambda}\frac{p^2}{E(M,p)}\left(1-\frac{M_{\sigma}^2-4M^2}{M_{\sigma}^2-4E(M,p)^2}\right)\left(1- n(p,\mu_r)-\bar{n}(p,\mu_r)\right)dp = 0.
\ea
\end{subequations}
\section{Numerical Results}
\label{sec-3}

In this section, we calculate the numerical results for the pole masses of the $\pi$ and $\sigma$ mesons. Firstly, by fitting the physical pion mass $M_{\pi} = 137$ MeV , decay constant $f_{\pi} = 93$ MeV and the quark condensate $\langle \bar{u}u\rangle =-(247)^3 \, \rm{MeV}^3$, we obtain the current mass of light quarks $m = 5.5$ MeV, the three momentum hard cutoff $\Lambda = 631$ MeV and the coupling constant $g = 5.074\time 10^{-6}$ MeV$^{-1}$ in the conventional NJL model\cite{Asakawa:1989bq}. And then, the new coupling constant G in the modified NJL model is given by
\ba
G = \frac{1+\frac{1}{N_c}}{1-\alpha+\frac{\alpha}{4N_c}}g.
\ea

\begin{figure}[t]
\centering
\includegraphics[width=0.8\textwidth]{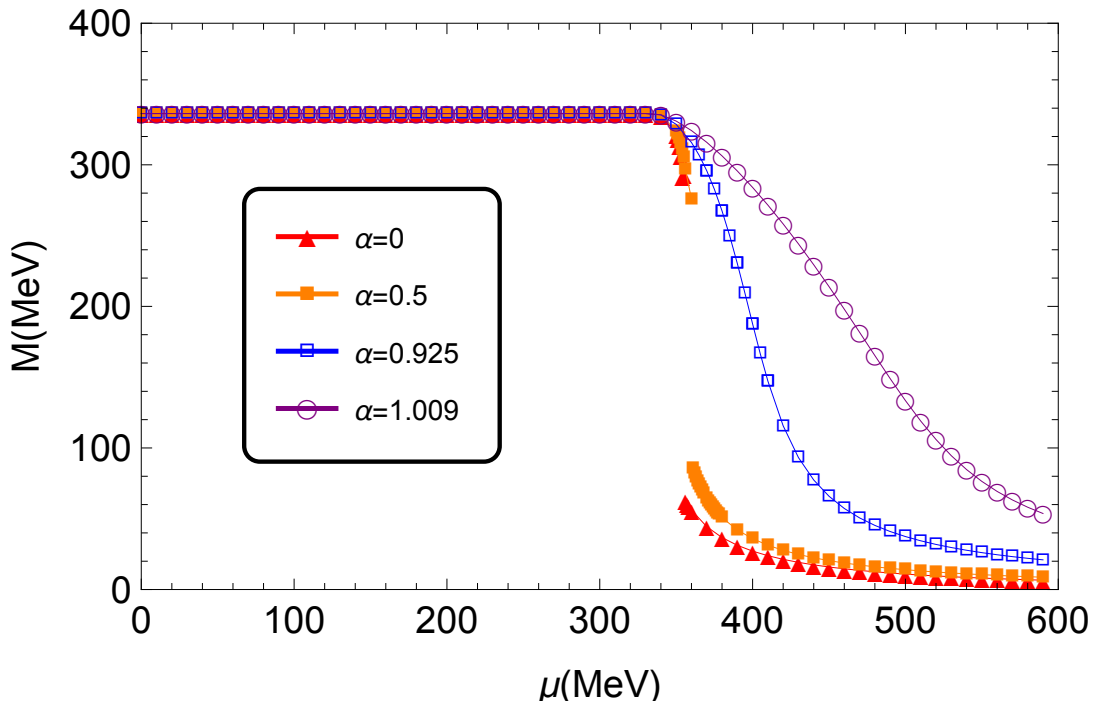}
\caption{The constituent qaurk mass $M$ as a function of the chemical potential $\mu$ at $T = 0$ and $\alpha~=~0,~0.5,~0.925~\text{and}~1.009$, respectively.}
\label{Fig1}
\end{figure}
\begin{figure}[t]
\centering
\subfigure[]{\includegraphics[width=0.45\textwidth]{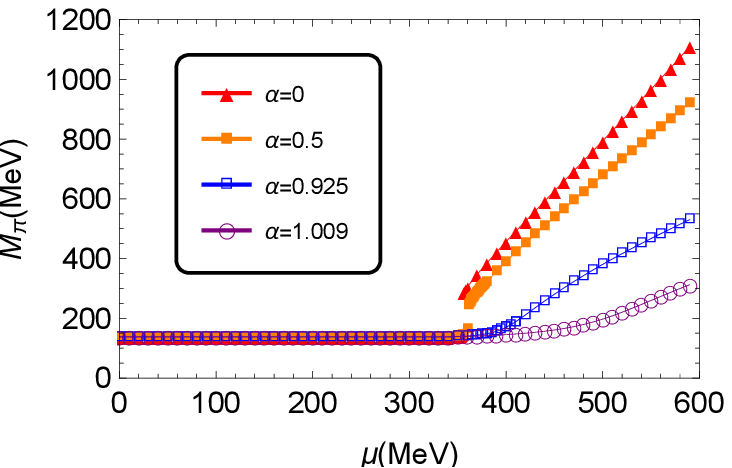}}
\subfigure[]{\includegraphics[width=0.45\textwidth]{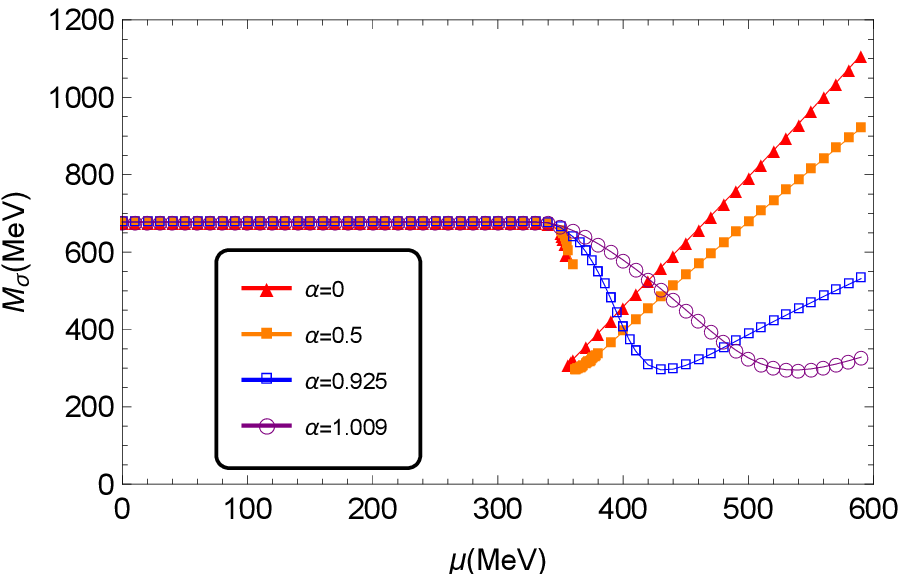}}
\caption{The pole mass of pion(left panel) and $\sigma$ meson(right panel) as a function of chemical potential at $T = 0$ and $\alpha~=~0,~0.5,~0.925~\text{and}~1.009$.}
\label{Fig2}
\end{figure}

\begin{figure}[t]
\centering
\includegraphics[width=0.4\textwidth]{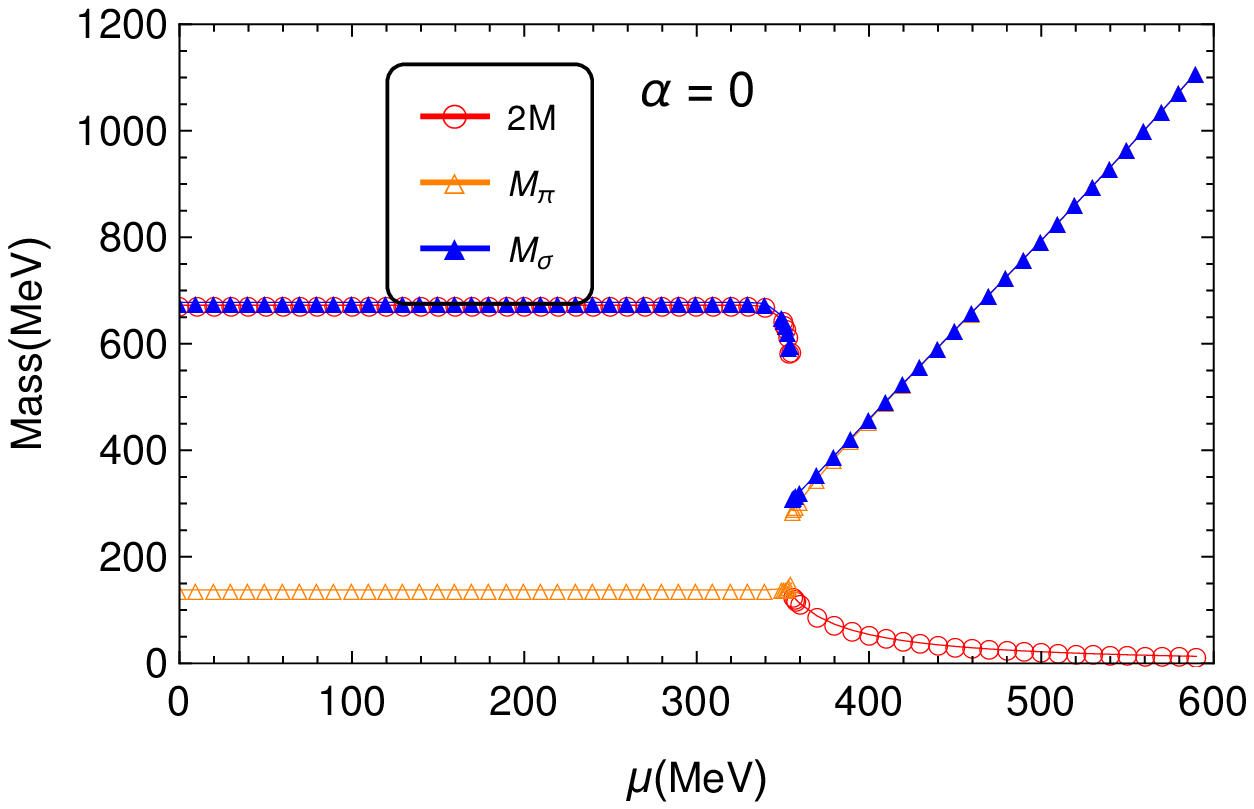}
\includegraphics[width=0.4\textwidth]{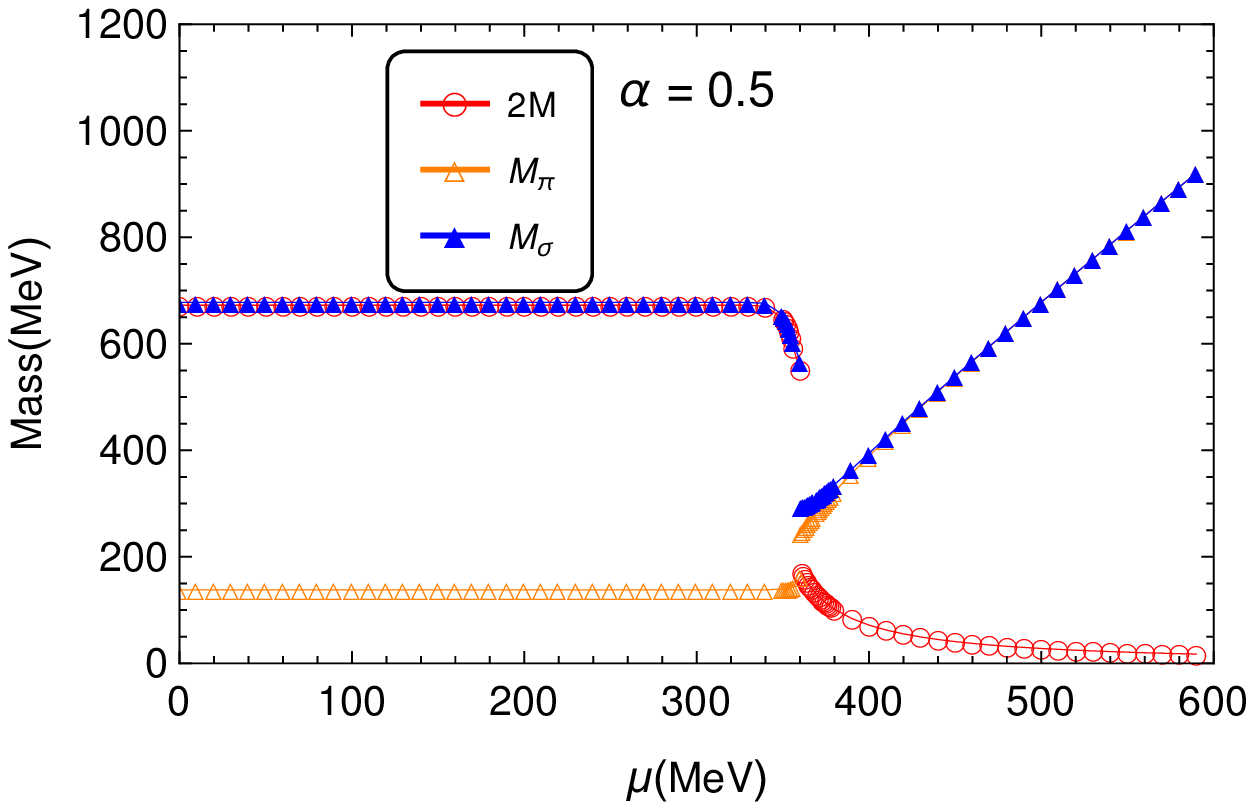}
\includegraphics[width=0.4\textwidth]{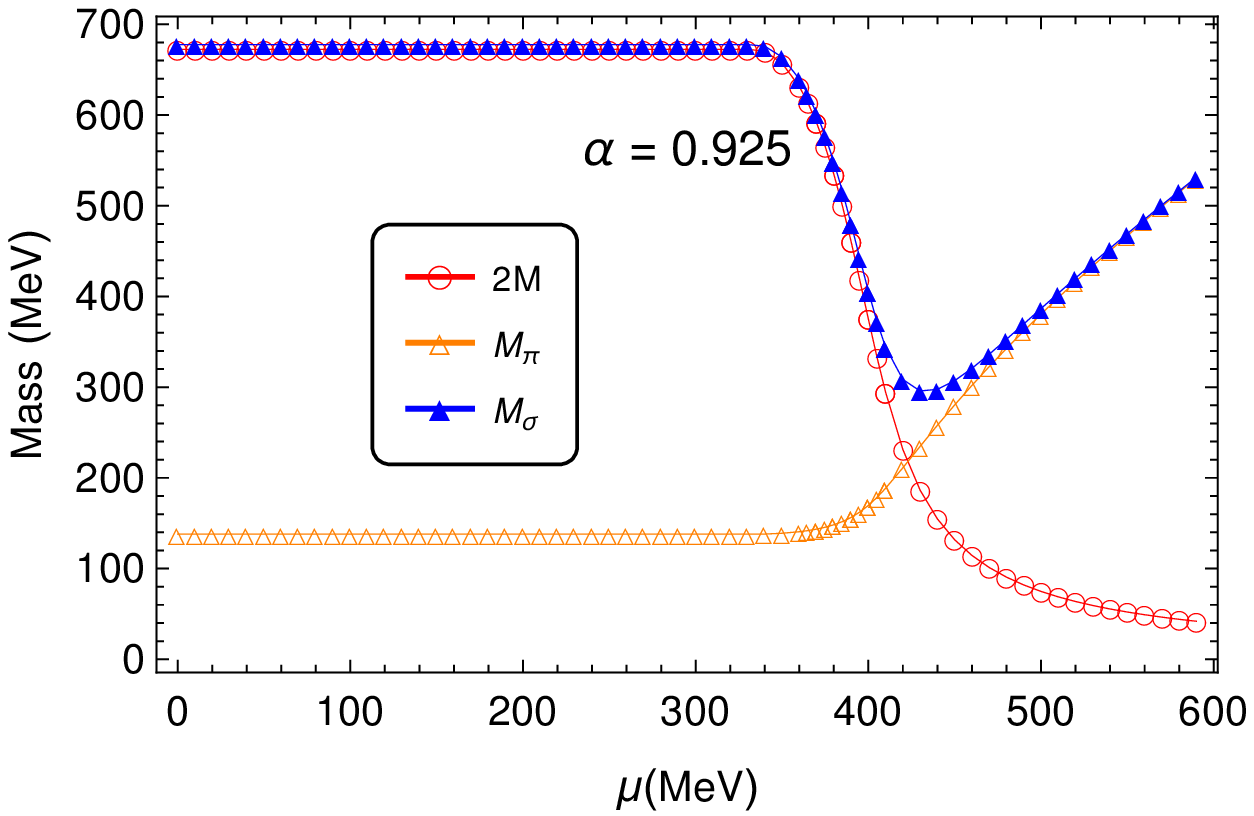}
\includegraphics[width=0.4\textwidth]{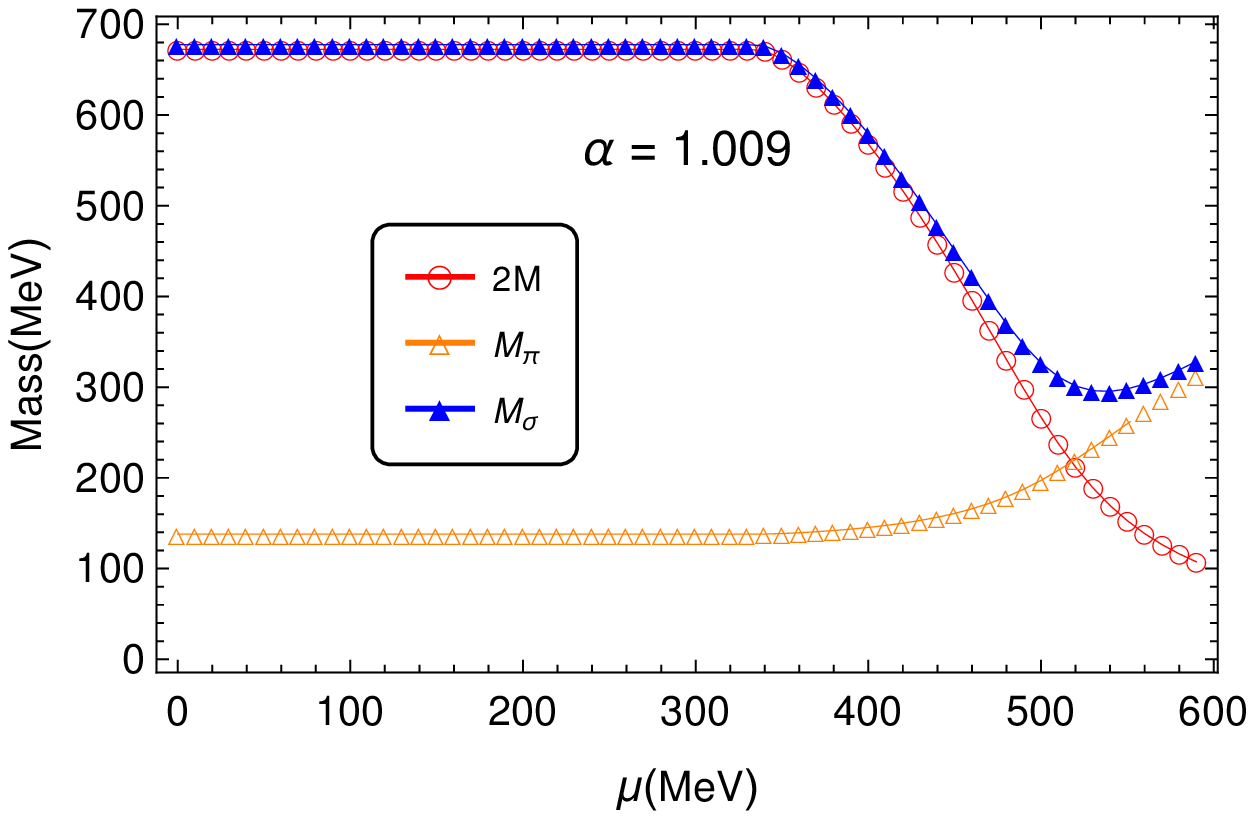}
\caption{The pole mass of $\sigma$ meson, pion and twice constituent quark mass as functions of chemical potential at $T = 0$ and $\alpha~=~0,~0.5,~0.925~\text{and}~1.009$.}
\label{Fig4}
\end{figure}

The constituent quark masses $M$ as a function of quark chemical potential $\mu$ at zero temperature but with some different weighting constants $\alpha$  are plotted in Fig.~\ref{Fig1}. And it is easy to find that, within our model parameters, the order of chiral phase transition changes from first order to crossover as $\alpha$ increases from zero: explicitly, they are first-order phase transitions at $\alpha = 0, 0.5$, while they are crossover transitions at $\alpha = 0.925, 1.009$. Hence, there must be a threshold value $\alpha_c$ between 0.5 and 0.925 where the termination of the first-order transition happens. And this has also been discussed in Ref.~\cite{Wang:2019uwl} by using different NJL parameters and will be further investigated in Ref.~\cite{Yu:2023njw} with the same parameters. Moreover, the (pseudo)critical chemical potential $\mu_c$ at zero temperature is found to be located at about 354, 360, 397 and 474 MeV, respectively, by analyzing the data in Fig.~\ref{Fig1} numerically for these four different values of $\alpha$ from 0 to 1.009. And another point worth mentioning is that, when $\mu \lesssim340$ MeV, the constituent quark mass $M$ remains constant for all different values of $\alpha$ at zero temperature. As for $\mu \gtrsim340$ MeV, we can see that the smaller $\alpha$, the more rapidly the constituent quark mass decreases. Therefore, the (approximate) chiral symmetry will be fully restored at a larger chemical potential when $\alpha$ becomes larger.

The pole masses of $\pi$ and $\sigma$ mesons as functions of quark chemical potential at zero temperature are given in Figs.~\ref{Fig2}, which obtained by Eqs.~(\ref{GapEq-pion}) and  (\ref{GapEq-sigma}) with different weighting constants $\alpha$. Note that, when $\alpha = 0$, our results coincide with the previous results in the conventional mean field approximation~\cite{Asakawa:1989bq}. Similar to the constituent quark mass, the meson masses keep constant as long as the chemical potential is smaller than $340$ MeV. When the chemical potential reaches $340$ MeV, the pion mass begins to increase with $\mu$, while the $\sigma$ meson mass decreases first and then increases. And in the region $\mu \gtrsim340$ MeV, the smaller $\alpha$ is, the faster the meson masses change. Also there is a jump at $\mu_c$ for small $\alpha$ where a first-order chiral phase transition happens.

The pole masses of $\pi$ and $\sigma$ mesons as well as twice the constituent quark mass for some selected $\alpha$ are plotted in Fig.~\ref{Fig4}. Note that the Mott chemical potential is defined by $M_{\pi}(\mu_{Mott}) = 2M(\mu_{Mott})$. The Mott transition is the signification where pions dissociate to the unbound resonance state. As discussed in Ref.~\cite{Mao:2019avr}, when the chiral phase transition is a smooth crossover, instead of the definition of the pseudo critical point, the Mott transition point is the better way to characterize the chiral crossover. In our case, when $\alpha$ is small, $\mu_{Mott}$ and $\mu_c$ are almost identical. When $\alpha$ is large, $\mu_{Mott}>\mu_c$, the differences are larger when $\alpha$ is larger ($\mu_{Mott} = 354,~360,~420~\text{and}~519$ MeV for $\alpha = 0,~0.5,~0.925~\text{and}~1.009$, respectively). The smoother the phase transition($\alpha$ is larger in our model), the larger the difference between $\mu_{Mott}$ and $\mu_c$. For $\sigma$ meson, the dissociation chemical potential is defined as $m_{\sigma}(\mu_d) = 2m_{\pi}(\mu_d)$. From our data, $\mu_d =354,~360,~402,~483$ MeV for $\alpha = 0,~0.5,~0.925~\text{and}~1.009$, respectively. The difference between $\mu_d$ and $\mu_c$ is also enhanced when $\alpha$ becomes larger, but the difference is tiny even when $\alpha$ is larger than 1. At very large chemical potential, the pole mass of $\sigma$ meson and pion are almost identical (they are the same in the chiral limit, while for massive light quark current mass $M_{\pi}^2 = M_{\sigma}^2 - 4 m^2 $). Of course, the larger the $\alpha$ is, the larger chemical potential is needed to reach such feature.

\begin{figure}[t]
\centering
\subfigure[]{\includegraphics[width=0.45\textwidth]{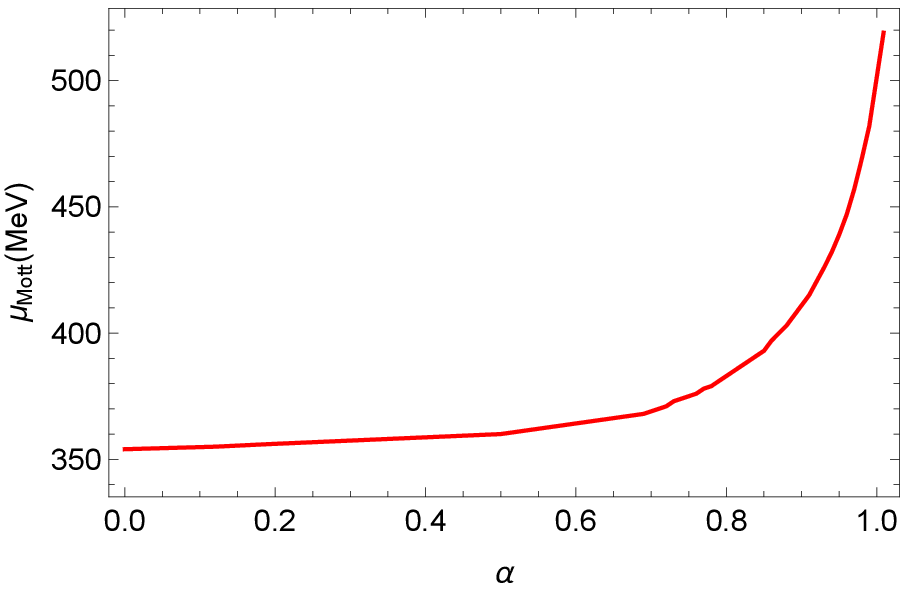}}
\hspace{0.01\textwidth}
\subfigure[]{\includegraphics[width=0.45\textwidth]{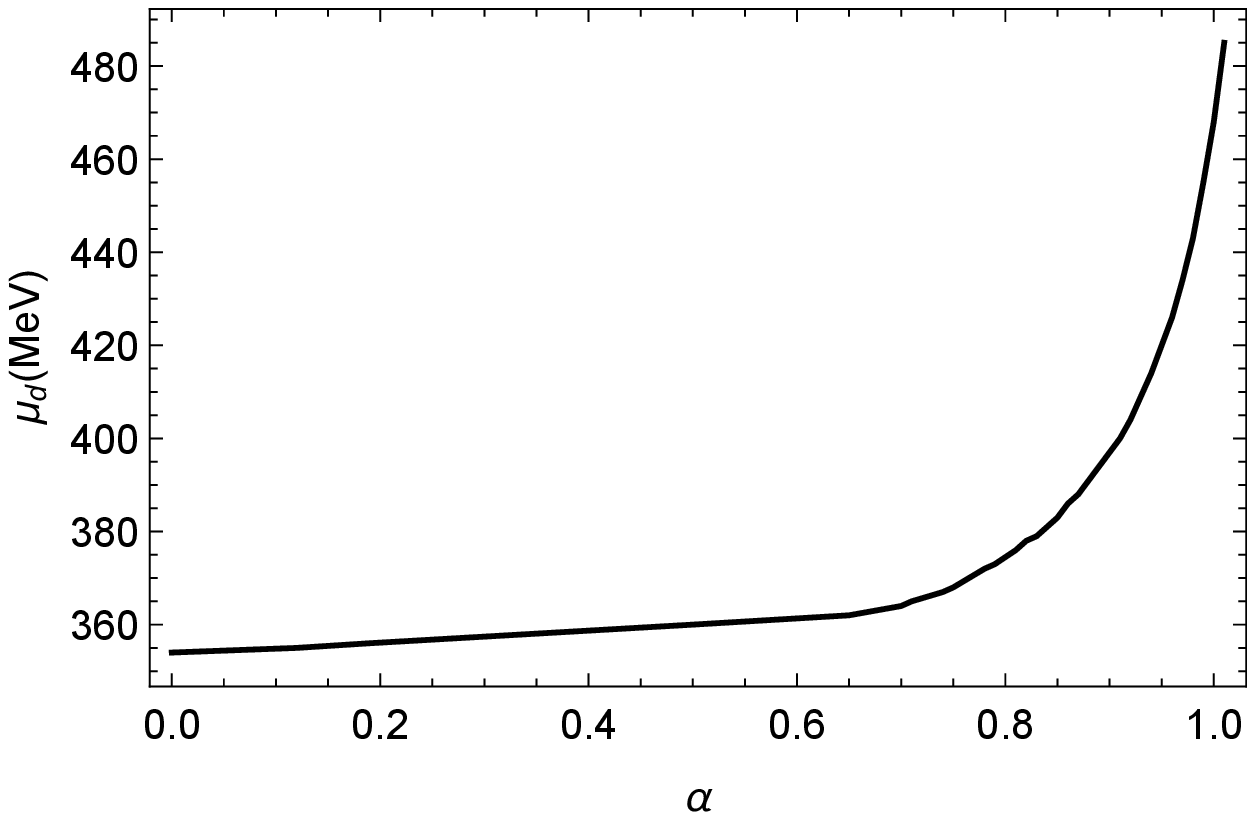}}
\caption{The Mott chemical potential(left panel) and the dissociation chemical potential(right panel) as a function of $\alpha$.}
\label{Fig5}
\end{figure}

To further clarify the impact of $\alpha$ on the Mott chemical potential and dissociation chemical potential, we have plotted them as functions of $\alpha$ in Fig.~\ref{Fig5}. The Mott chemical potential and the dissociation chemical potential are shown in the left and right panels, respectively. Both functions display an exponential increase with $\alpha$. Additionally, the Mott chemical potential is higher than the dissociation chemical potential for large values of $\alpha$. These results support the discussion presented in the previous paragraph.

\begin{figure}[t]
\centering
\subfigure[]{\includegraphics[width=0.45\textwidth]{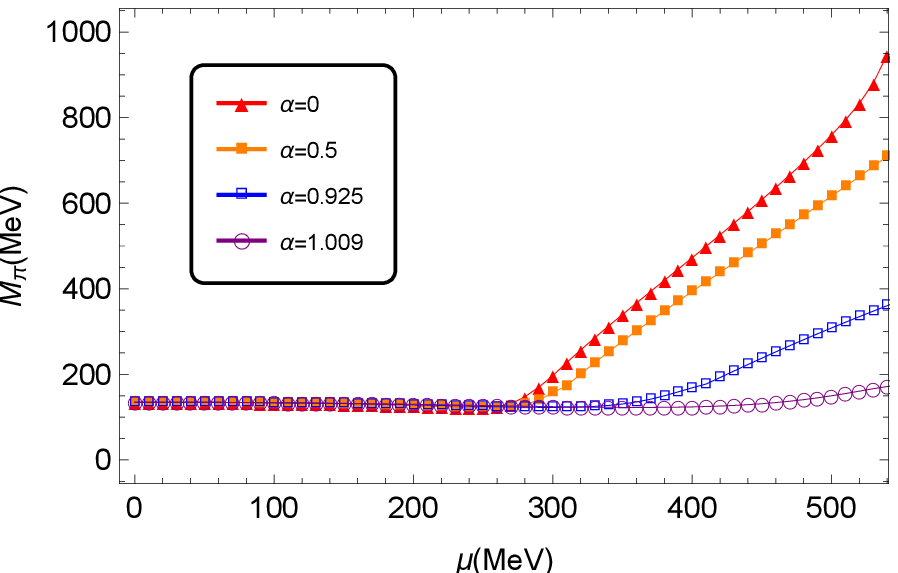}}
\hspace{0.01\textwidth}
\subfigure[]{\includegraphics[width=0.45\textwidth]{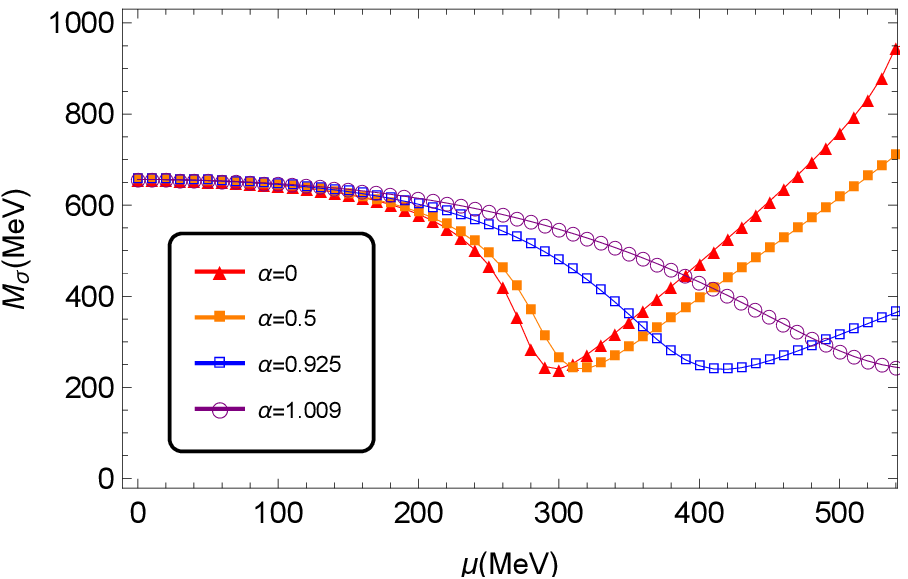}}
\caption{The pole mass of pion(left panel) and $\sigma$ meson(right panel) as a function of chemical potential at $T = 100$~MeV and $\alpha~=~0,~0.5,~0.925~\text{and}~1.009$.}
\label{Fig6}
\end{figure}

Additionally, for finite temperature case, the only differences with zero temperature case is that the first-order phase transition regime becomes smaller as the temperature increases. The discontinuity in the mesons mass disappears for all values of $\alpha$ above a certain temperature. We have demonstrated this by plotting the pion and $\sigma$ meson masses at a temperature of $T = 100$~MeV in Fig.~\ref{Fig6}. Our model predicts that at this temperature, the phase transition is a crossover for all values of $\alpha$. The mesons mass do not exhibit a ``jump" as a function of chemical potential, and the rest of the physical behaviors coincide with our zero-temperature results.

\section{Discussion and conclusion}
\label{sec-4}
In this study, we investigate the pole masses of pion and $\sigma$ mesons as functions of the chemical potential $\mu$ in the context of the Nambu-Jona-Lasinio (NJL) model using a new self-consistent means field approximation method. We find that the mass spectrum of mesons can provide insight into the chiral phase transition of quantum chromodynamics (QCD) matter. In particular, we observe that at $T = 0$, the chiral phase transition can be either first order or a crossover depending on the value of $\alpha$. Additionally, the difference between the Mott transition and chiral phase transition becomes more significant as $\alpha$ increases, and we find that the temperature can help to restore chiral symmetry, leading to a continuous mass spectrum for any value of $\alpha$.
In the high density environment such as in the core of compact stars, the chemical potential of quark matter should reach several hundreds of MeV. The weighting constant $\alpha$ is an important feature to study the dense matter, in addition to the indirect measurements which constrain the equation of state (EOS) of neutron stars. Our results of the $\pi$ and $\sigma$ mesons show the significant differences in meson properties with different $\alpha$ values.

We suggest that the meson properties in collisions with large chemical potential or in compact stars could be a good way to study the chiral phase transition of quantum chromodynamics (QCD) matter. Furthermore, the meson mass differences for different values of $\alpha$ can be measured by studying the decay channels of the $\pi^0$ and $\sigma$ mesons. The electromagnetic probes (photons and dilepton) have a long mean-free-path and minimal interactions with the medium in collision experiments, making them excellent choices to study the meson properties, e.g., $\pi^{0} \rightarrow \gamma+ \gamma$ and   $\pi^0\rightarrow \gamma + \gamma\rightarrow e^{+}+e^{-}+\gamma$. In the future, electromagnetic probes may be used in high chemical potential collision experiments to study the meson properties. Additionally, if the "sign problem" can be solved, lattice calculations could provide a more straightforward way to determine the value of $\alpha$ in the NJL model.

\acknowledgements
The work of X.W. is supported by the start-up funding No.~4111190010 of Jiangsu University and NSFC under Grant No.~12147103.

\end{document}